\documentclass[aps,prd,preprint,groupedaddress]{revtex4-1}
\usepackage{graphicx}
\usepackage{epsfig}
\usepackage[caption=false]{subfig}

\begin{document}


\title{Jet fragmentation as a tool to explore double parton scattering using Z-boson + jets processes at the LHC}



\author{R.~Kumar}
\email[]{raman\_phy@auts.ac.in}
\affiliation{Akal University, Talwandi Sabo, INDIA}

\author{M.~Bansal}%
\affiliation{D.A.V. College, Sector 10, Chandigarh, INDIA}

\author{S.~Bansal}
\email[]{Sunil.Bansal@cern.ch}
\affiliation{Panjab University, Chandigarh, INDIA}


\date{\today}

\begin{abstract}

The Large Hadron Collider witnesses the highest ever production cross-section of double parton 
scattering processes. The production of a Z-boson along with two jets 
from double parton scattering provides a unique opportunity to explore the kinematics of 
double parton scattering processes and their dependence on the scale of the second interaction. 
The experimental measurement of this process is largely contaminated by Z + jets production 
from single parton scattering. In this paper, fragmentation properties of a jet are 
explored to check their sensitivity towards double parton scattering. The present study 
is performed using simulated Z + jets events, produced with \textsc{madgraph} and 
\textsc{powheg} Monte-Carlo event generators, hadronized and parton showered using 
\textsc{pythia}8. 
The effect of different hadronization model on the discrimination based on the fragmentation 
properties of a jet is also investigated by using events simulated with \textsc{herwig}++.
It is observed that discrimination based on the fragmentation 
properties of a jet can significantly suppress the background from single parton 
scattering, which results into 40--50\% gain in the contribution of double 
parton scattering. 
\end{abstract}


\maketitle

\section{Introduction}

The large collision energy and involved parton densities in a proton-proton (pp) collision at Large Hadron 
Collider (LHC) lead to significant increase in probability of more than one parton-parton scattering 
in the same pp collision. The additional scatterings, along with primary hard parton-parton scattering, 
are called multiple parton interactions (MPI)~\cite{Sjostrand:1986ep}. Usually, MPI produce particles with 
relatively small transverse momenta ($p_{\rm T}$), but there is a possibility of producing particles with 
large $p_{\rm T}$ or mass, such as  jets and/or vector (W/Z) bosons. 
The production of such particles from at least 
two parton-parton scatterings is referred to as double parton scattering (DPS). The study of DPS 
provides vital information on the parton-parton correlations and parton distributions in a 
hadron~\cite{Diehl:2011yj}. In addition, DPS also 
constitutes as a background to new physics searches~\cite{Hussein:2006xr,SUSY}. A broad range of measurements 
is available for DPS processes at different collision energies, performed using different final states, $e.g.$, 
W/Z + jets, photon + jets, 4-jets, diboson processes~\cite{jetua,jetafs,photon3jetd0,photon3jetcdf,4jetcdf,w2jetcms,w2jetatlas,photon3jetcms,dpswwcms8,Aaboud:2016dea}, $etc$.

The production of Z + 2-jets from DPS is important because of large production cross-section and it also
provides the opportunity to study the dependence on the scale of the second hard interaction, as predicted 
by theoretical quantum chromodynamics (QCD) models~\cite{Blok:2017alw, Snigirev:2010tk}. The experimental 
measurements of these DPS processes are dominated by production of Z + jets from single parton scattering 
(SPS). In order to increase DPS sensitivity and study its kinematic properties, it is essential to devise 
methods to control SPS background with minimal effect on the DPS. In the existing experimental as well as
theoretical studies, DPS and SPS processes are distinguished using the kinematical and correlation 
properties of Z-boson and jets~\cite{Blok:2015afa, Maina:2010vh, Cao:2017bcb, w2jetcms, w2jetatlas}. This 
paper presents the studies, using Z + jets process, to demonstrate that fragmentation properties of a jet 
can be effective in suppressing the SPS background.
 
The outline of the paper is as follows. The simulation of the Z + jets events and selection criteria is 
discussed in Section~\ref{sec:EGnSC}. In Section~\ref{sec:methodology}, fragmentation properties of 
a jet and related variables are discussed. The results of the present study are discussed in the 
Section~\ref{sec:results} and finally Section~\ref{sec:summary} summarizes the paper.

\section{Event generation and selection criteria}\label{sec:EGnSC} 

\subsection{Event generation}
The present study requires events simulated with Monte-Carlo event generators, which are able to produce 
sufficient number of jets in association with Z-boson. The event generators 
\textsc{madgraph}~\cite{Alwall:2011,Maltoni:2003} and \textsc{powheg}~\cite{Frixione:2007,PowhegW2J}
are used to simulate Z + jets events in the present study. \textsc{madgraph} is a tree-level matrix  
element event generator, which is able to produce up to 4 jets as per matrix element leading order (LO) 
calculations. The distributions of Z + jets, produced by LHC, are well described by 
\textsc{madgraph}~\cite{Khachatryan:2016crw}. \textsc{powheg} produces Z + 2-jets events up to 
next-to-leading order (NLO) using `Multi-scale improved NLO'' (MiNLO) 
method~\cite{MINLO, Campbell:2013vha}, which is also capable to describe well the jets production 
associated with Z-boson. These events are further hadronized and parton showered with 
\textsc{pythia}8~\cite{Sjostrand:2007gs}. MPI are included while parton showering with 
\textsc{pythia}8~\cite{Corke:2009pm}.

The production of Z + 2-jets events from DPS processes is simulated with \textsc{pythia}8 event 
generator, which is configurable to specify process involved and interaction scale independently 
for the two parton-parton scatterings. In the present analysis, first parton-parton scattering 
is configured to produce Z-boson, whereas two jets are produced from the second parton-parton 
scattering. 

All these processes are produced in pp collisions assuming a center-of-mass energy of 13 TeV and using ATLAS 
A14 tune~\cite{ATLAS:2012uec} with NNPDF2.3LO set of parton distribution functions (PDF) for the simulation 
of MPI model. This version of MPI tune is derived by fitting the underlying event data collected at the LHC.

Dijet QCD events are  simulated using \textsc{pythia8} with A14 tune. 
These processes are also simulated using \textsc{herwig}++ (version 2.7.0)~\cite{Bahr:2008pv} and CUETHppS1 tune~\cite{Khachatryan:2015pea}.

\subsection{Event selection}
In these simulated events,  Z-boson candidates are identified using four-momenta of the muons coming from the 
decay of Z-boson. An event is considered to have a Z-candidate if it satisfies the following conditions:
\begin{itemize}
 \item
   presence of two muons with $p_{\rm T}$ larger than 20 GeV/$c$ and absolute pseudorapidity ($\eta$) less than 2.5.
\item
   Invariant mass of the two selected muons is required to be in range of 60$-$120 GeV/${c^{\rm 2}}$.
 \end{itemize}
This selection criteria is motivated from the trigger and the background constraints for the measurements 
involving Z-boson at the LHC experiments. 
These events, having a Z candidate, are required to have at least two jets with a minimal $p_{\rm T}$ of 20 GeV/$c$ and 
$|\eta|<$ 2.5. The jets are clustered using the anti-$k_T$ algorithm~\cite{Cacciari:2008gp} with the radius 
parameter equal to 0.5 using the \textsc{FastJet} software package~\cite{Cacciari:2011ma}.


\section{Methodology}\label{sec:methodology}

\begin{figure}[htbp]
\begin{center}
\includegraphics[width=0.5\textwidth]{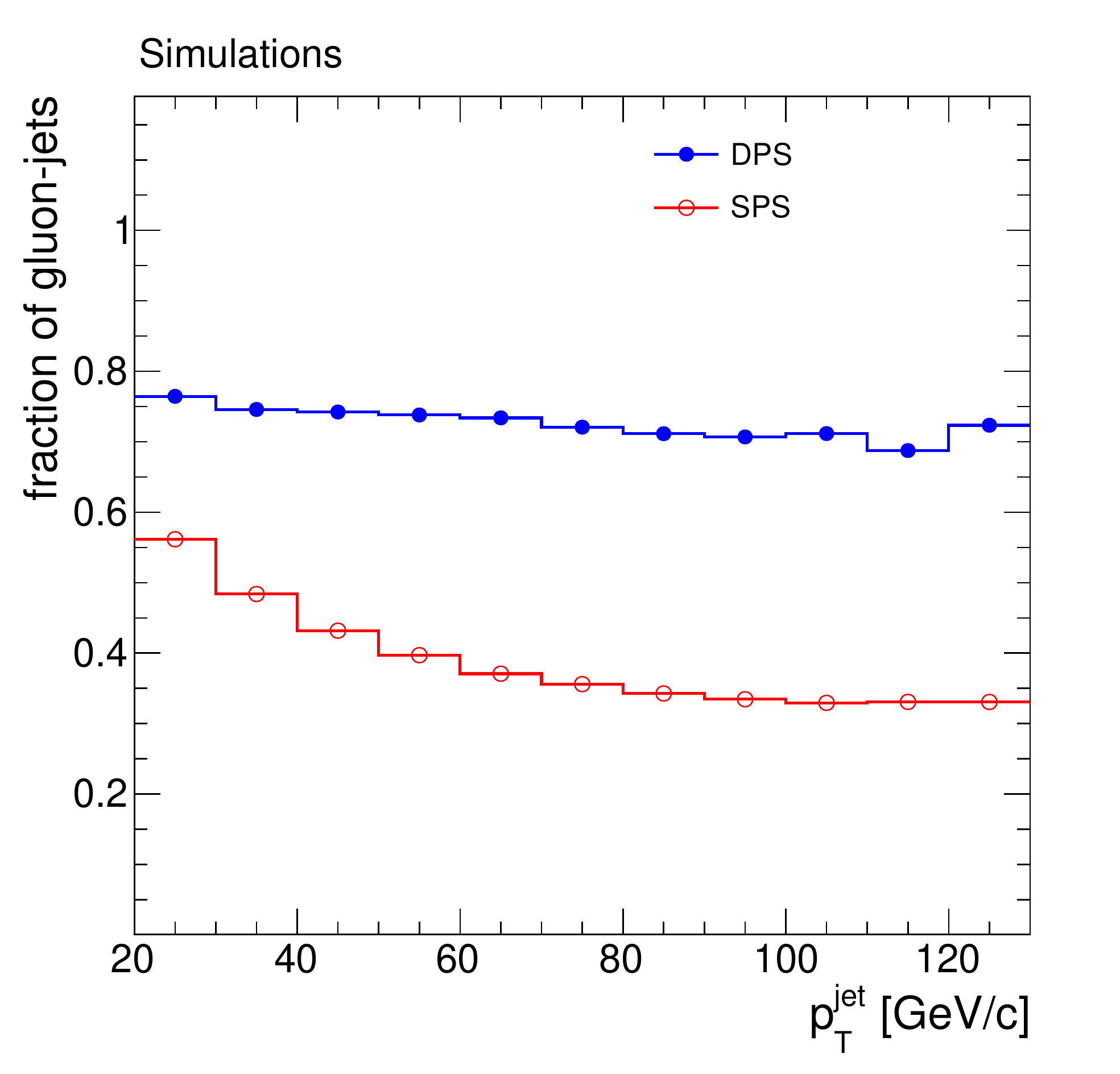}

\end{center}
\caption {{The fraction of gluon-initiated jets as a function of the jet $p_{\rm T}$ in the simulated 
DPS (blue solid circle markers) and SPS (red hollow circle markers) events.}} 
\label{qg-ratio}
\end{figure}

The production of Z + jets events from DPS includes 2-jets production from the second interaction, which is 
dominated with the jets initiated by gluons. Whereas, the jets produced in association with Z-boson 
via SPS are expected to be  predominately initiated by quarks. 
Figure~\ref{qg-ratio} shows the fraction of gluon-initiated jets, in Z + jets events produced via DPS 
and SPS, as a function of the jet $p_{\rm T}$. A jet is tagged as gluon-initiated or quark-initiated 
by matching the jet with partons in $\eta \times \phi$ space. A distance parameter $\Delta \rm R$ 
is defined as:

\begin{equation}
  \Delta R = \sqrt{(\eta_{\rm jet }- \eta_{\rm parton})^{2} + (\phi_{\rm jet} - \phi_{\rm parton})^{2}}.
\end{equation}
A pair of jet-parton is considered to be matched if $\Delta \rm R$ is smaller than 0.3. It is observed 
that $\approx$ 75\% of jets in the DPS sample are gluon-initiated whereas in SPS sample the fraction of 
gluon-initiated jets is relatively small ($\approx$ 45\%). Thus, contribution of DPS events can be 
increased if flavor of a jet can be identified and only events with gluon-initiated 
jets are considered.

It is well established that the fragmentation properties of the quarks and the gluons are different from 
each other~\cite{OPAL1993:qg1,OPAL:1995ab,Abreu:1995hp,Buskulic:1995sw,Gallicchio:2011xq,Aad:2014gea,CMS:2013kfa}.
In particular, the jets initiated by the gluons show significantly different behavior with respect to those 
initiated by light-flavor quarks (u, d, s). There are number of 
observables~\cite{Gallicchio:2011xq,Aad:2014gea,CMS:2013kfa,Cornelis:2014ima} which can be constructed using 
intrinsic properties of a jet. 
The study presented in this article is aimed to emphasize on use of intrinsic properties of jets for SPS background suppression rather than to construct the best quark-gluon discriminator.
Therefore, only a certain number of observables are considered in the analysis as discussed below:

\begin{itemize}
 \item {\bf Jet size ($\sigma^{\rm jet}$)}: A jet, with conical structure, is approximated by an ellipse when 
 projected in $\eta \times \phi$ space. Two principal axes of the ellipse are used which are represented as 
 major axis  ($\sigma_{1}^{\rm jet}$) and minor axis ($\sigma_{2}^{\rm jet}$) of the jet cone.
 \item {\bf Jet constituents multiplicity ($N_{p}^{\rm jet}$)}, which is the total number  of particles  
 (charged or neutral) within the jet.
 \item {\bf Jet fragmentation function ($p_{\rm T}^{\rm jet}{D}$)}, which represents the $p_{\rm T}$-distribution  
 among the constituents of a jet. It is defined as:
 
\begin{equation}
     p_{\rm T}^{\rm jet}{D} = \frac{ \sqrt{\sum_{i} p_{\rm T,i}^{2}}}{\sum_{i} p_{\rm T,i}},
    \end{equation}

where sum extends over the jet constituents and $p_{\rm T,i}$ represents the $p_{\rm T}$ of $i^{th}$  constituent. 
The value of $p_{\rm T}^{\rm jet}{D}$ for a jet will be equal to one if there is only one constituent for that particular 
jet. The jets with infinite number of jet constituents will have value of $p_{\rm T}^{\rm jet}{D}$ approaching 
towards zero. 
\end{itemize}

\begin{figure}[htbp]                                                                  
\begin{center}
\dimendef\prevdepth=0
\subfloat[]{\includegraphics[width=0.45\textwidth]{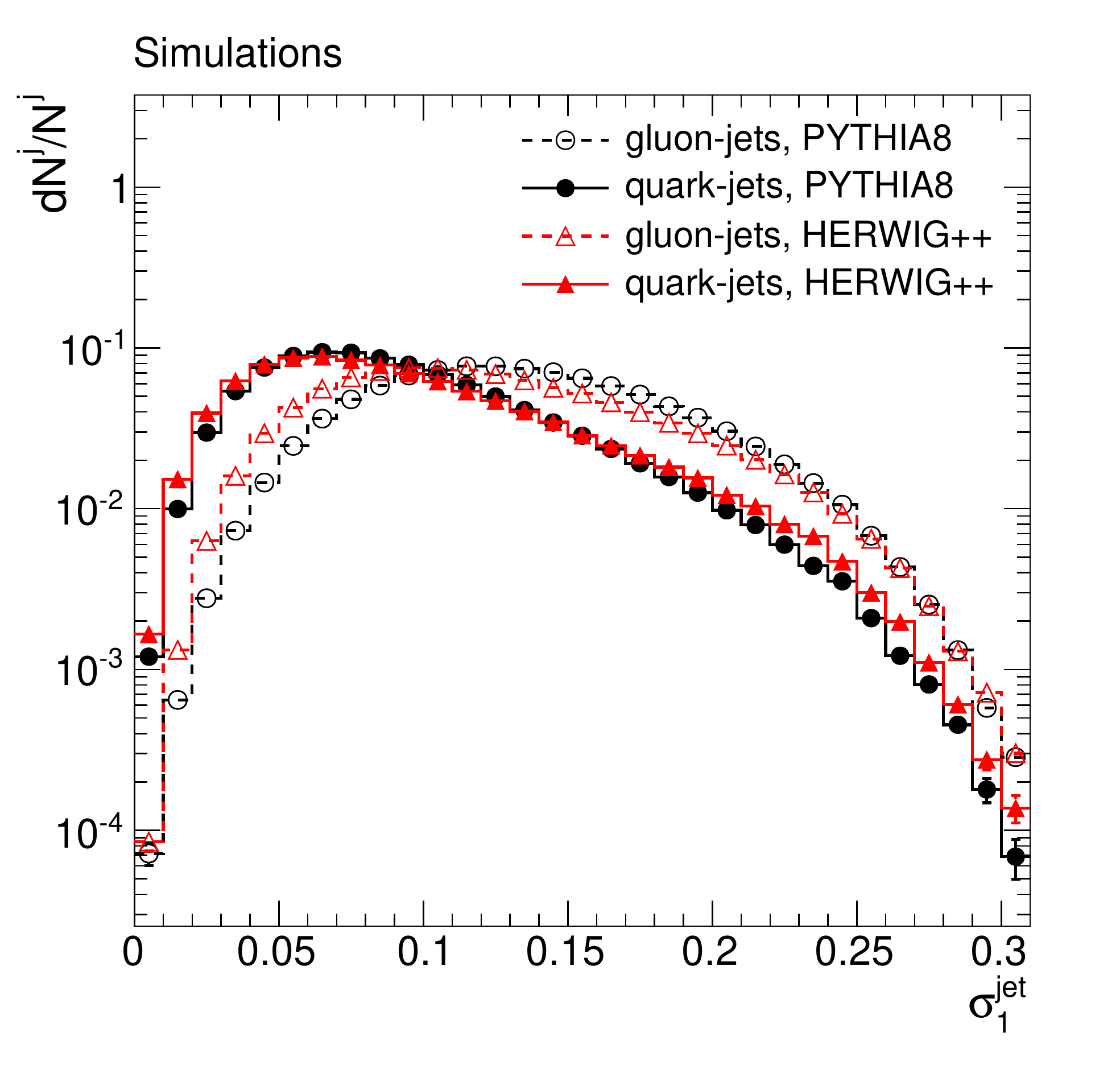}}
\subfloat[]{\includegraphics[width=0.45\textwidth]{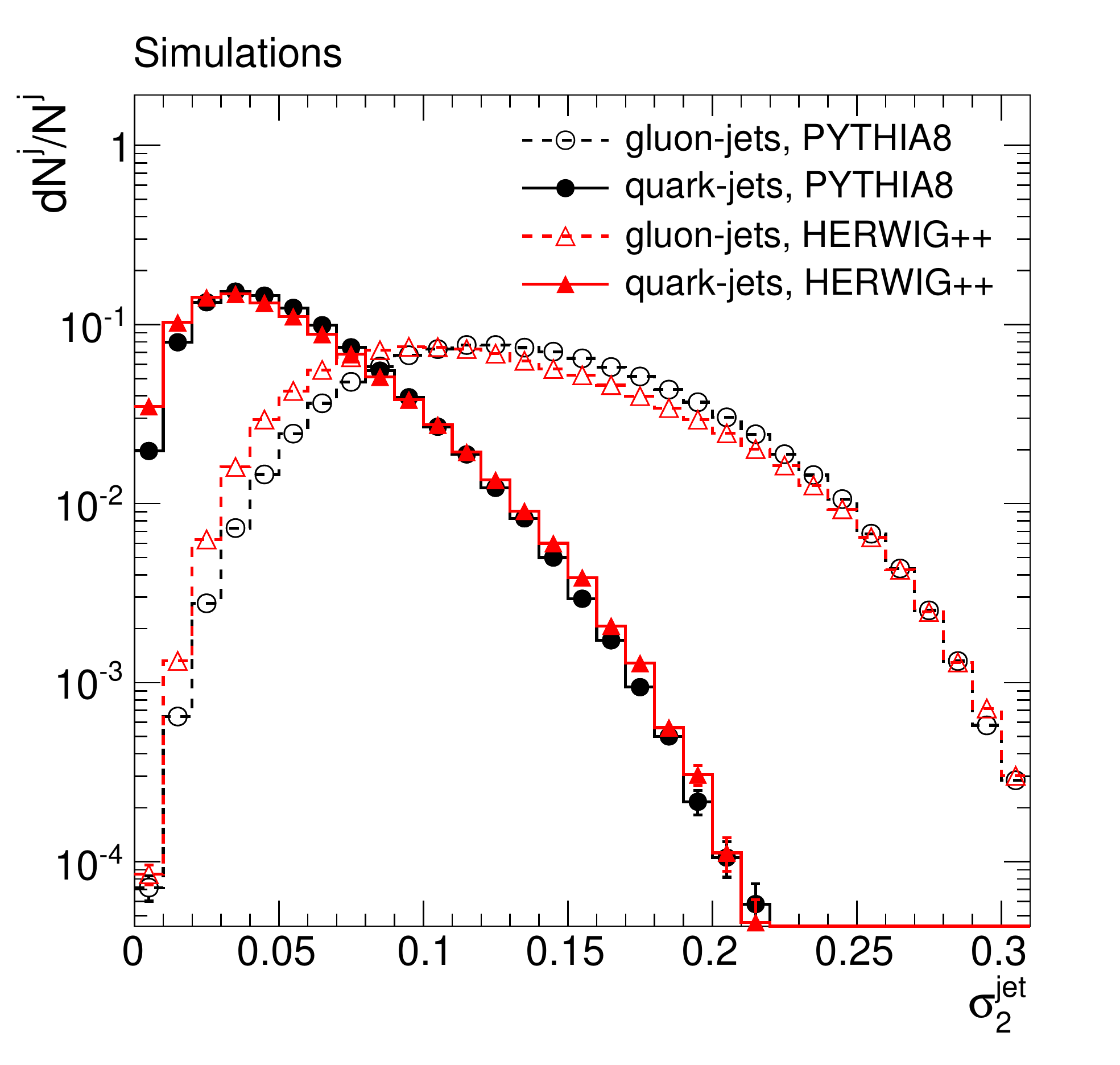}}\\
\subfloat[]{\includegraphics[width=0.45\textwidth]{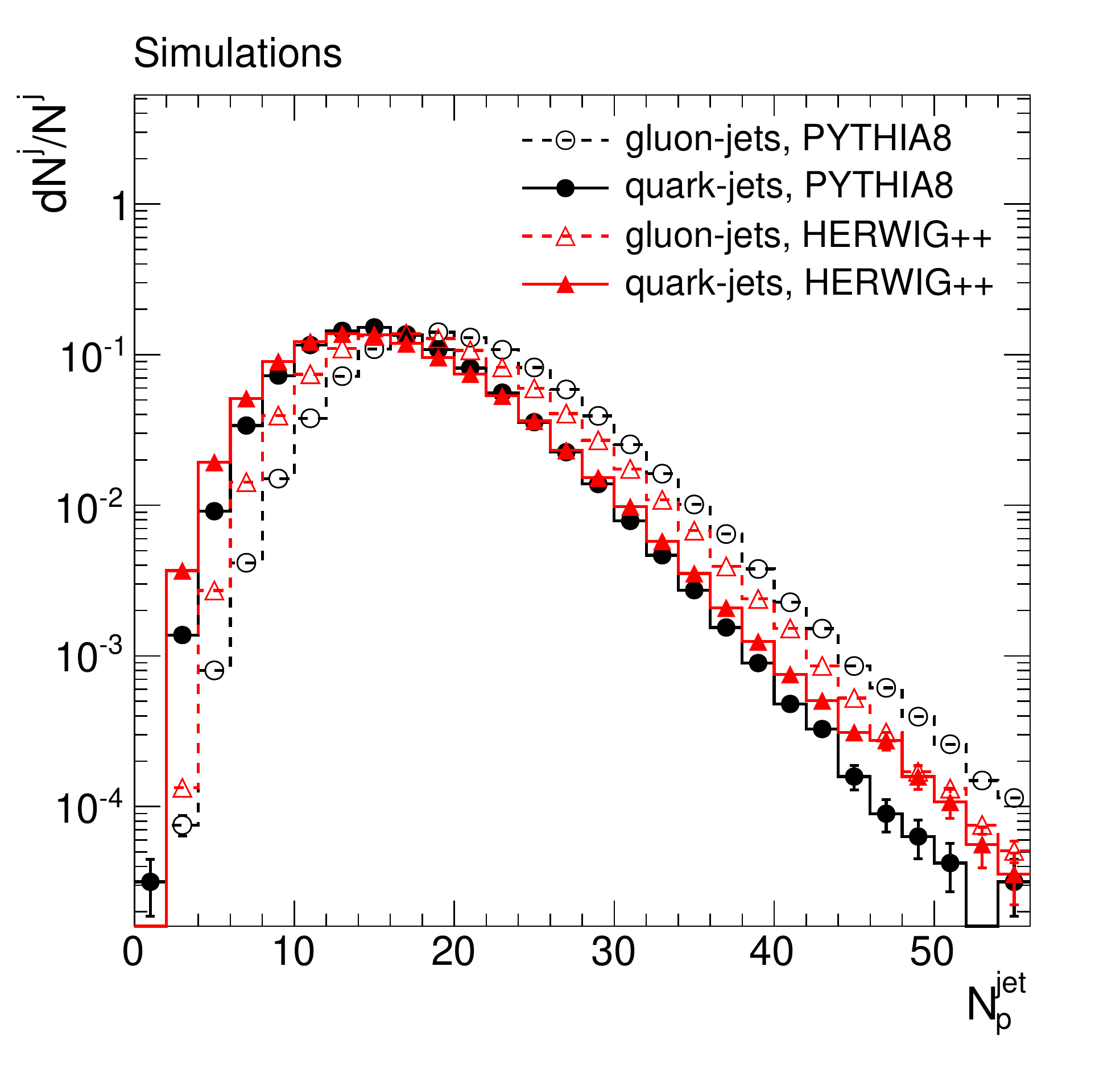}}
\subfloat[]{\includegraphics[width=0.45\textwidth]{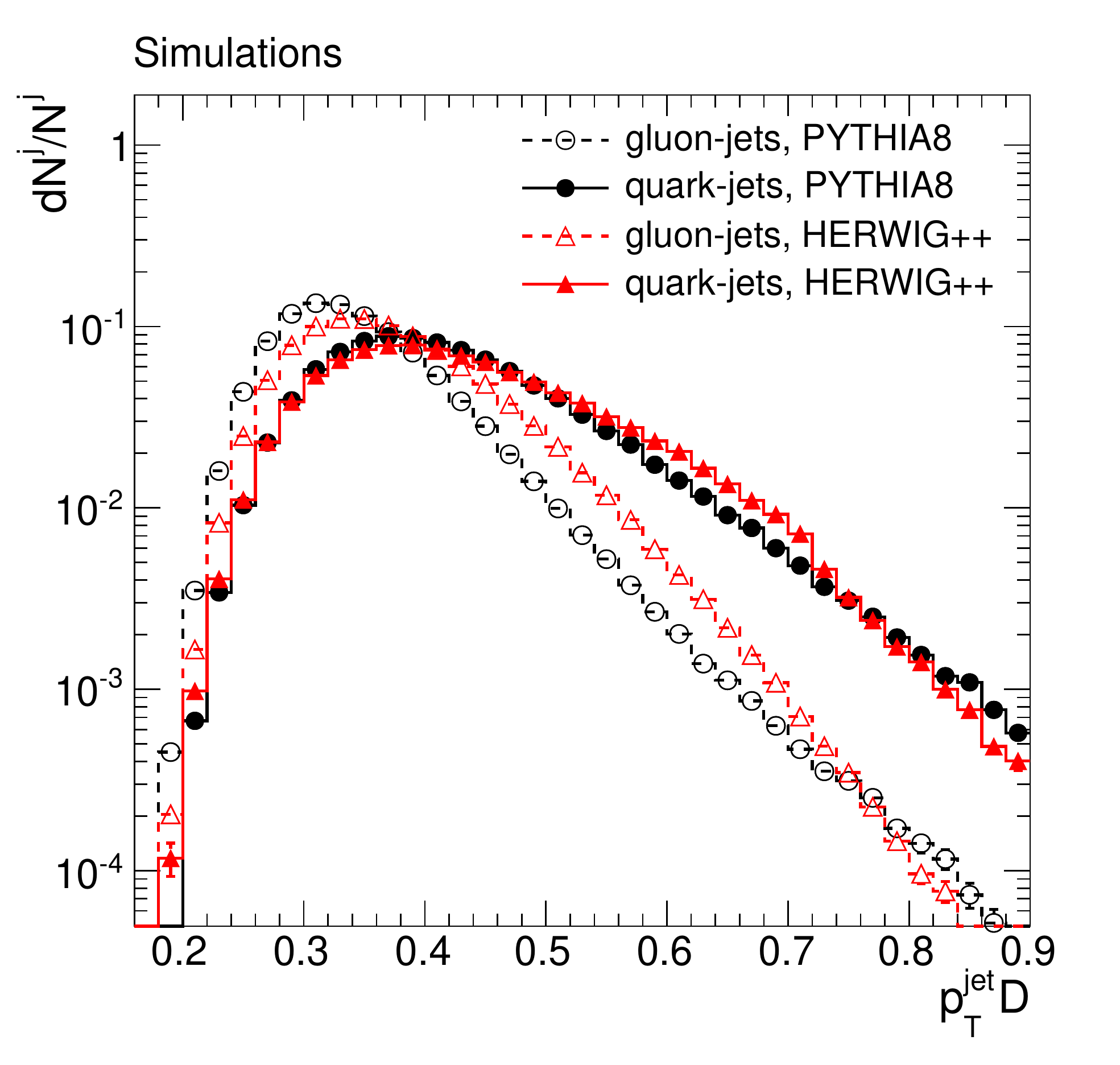}}

\caption {{The distributions of discriminating variables: 
(a) major axis size of jet cone ($\sigma_{1}^{\rm jet}$), (b) minor axis size of jet cone ($\sigma_{2}^{\rm jet}$), 
(c) jet constituents multiplicity ($N_{p}^{\rm jet}$), and (d) jet fragmentation function ($p_{\rm T}^{\rm jet}{D}$)
are compared for gluon-initiated jets (hollow markers) and quark-initiated jets (solid markers) events hadronized and parton showered with \textsc{pythia}8 (black colored markers) and \textsc{herwig}++ (with red colored markers).}} 
\label{comp_inpVarBDT2}
\end{center}
\end{figure}

More details of these observables can be found in Ref.~\cite{CMS:2013kfa,Cornelis:2014ima}. The distributions 
of these discriminating variables for gluon-initiated jets and quark-initiated jets are shown in 
Fig.~\ref{comp_inpVarBDT2}. These distributions are constructed using dijet events simulated with \textsc{pythia}8 and as well as using \textsc{herwig}++. 
The gluon-initiated and quark-initiated jets are identified with jet-parton matching in $\eta \times \phi$ space as described earlier. 
It is clear from these distributions that gluon-initiated jets are broader than quark-initiated jets and have larger number of particles. 
The constituents of the quark-initiated jets are hard as compared to the constituents of the gluon-initiated jets. 
Therefore, the quark-initiated jets have a harder fragmentation function as compared to gluon-initiated ones. 
Thus, for these observables a clear distinction between gluon- and quark-initiated jets is visible. 
The effectiveness of the usage of these variables is investigated with cut-based and multivariate analysis (MVA) methods.


\section{Results and Discussions}\label{sec:results}
In Z + jets events simulated with \textsc{madgraph} + \textsc{pythia}8, two jets are required along with a 
Z-boson as per the kinematic selection criteria mentioned in Section~\ref{sec:EGnSC}. The selected Z + 2-jets 
events contains the contribution from DPS as well as SPS processes. A selected event is considered to be 
produced by DPS, if there are two MPI partons present within the acceptance, otherwise event is 
considered as SPS background~\cite{w2jetcms}. The fraction of  DPS processes contributing in 
selected Z + 2-jets sample is about 0.075 which is consistent with the previous studies~\cite{w2jetcms}. 
The effect of the observables mentioned in Section~\ref{sec:methodology}, which are sensitive to 
quark-gluon discrimination, is evaluated by calculating the gain in DPS fraction after requiring selected 
two jets to be gluon-initiated. 

In cut-based method, the jets are considered to be gluon-initiated if two selected jets satisfy the 
conditions listed in Table~\ref{tab:cut}. These cuts are optimized by maximizing the figure of merit defined 
as $S/\sqrt{S+B}$, where $S$ represents gluon-initiated jets and $B$ represents quark-initiated jets.
This set of selection criteria selects $\approx$ 82\% of gluon-initiated 
jets with 54\% rejection of quark-initiated jets.
When both of the jets are required to be gluon initiated, DPS fraction comes out to be 0.106 which is 41\% larger as compared to selection when fragmentation properties of the jets were not used.

\begin{table}
\caption{\label{tab:cut}Conditions on observables for selection of gluon-initiated jets in cut-based analysis}
\begin{tabular}{c|c}
~~Observable~~ & ~~condition~~ \\
\hline
$\sigma^{\rm jet}_{1}$ & $>$~0.04 \\
\hline
$\sigma^{\rm jet}_{2}$ & $>$~0.02 \\
\hline
$N_{p}^{\rm jet}$ & $>$~12.0 \\
\hline
$p_{\rm T}^{\rm jet}{D}$ & $<$~0.49 \\

\end{tabular}
\end{table}

\begin{figure}[htbp]                                                                  
\begin{center}
\dimendef\prevdepth=0
\includegraphics[width=0.5\textwidth]{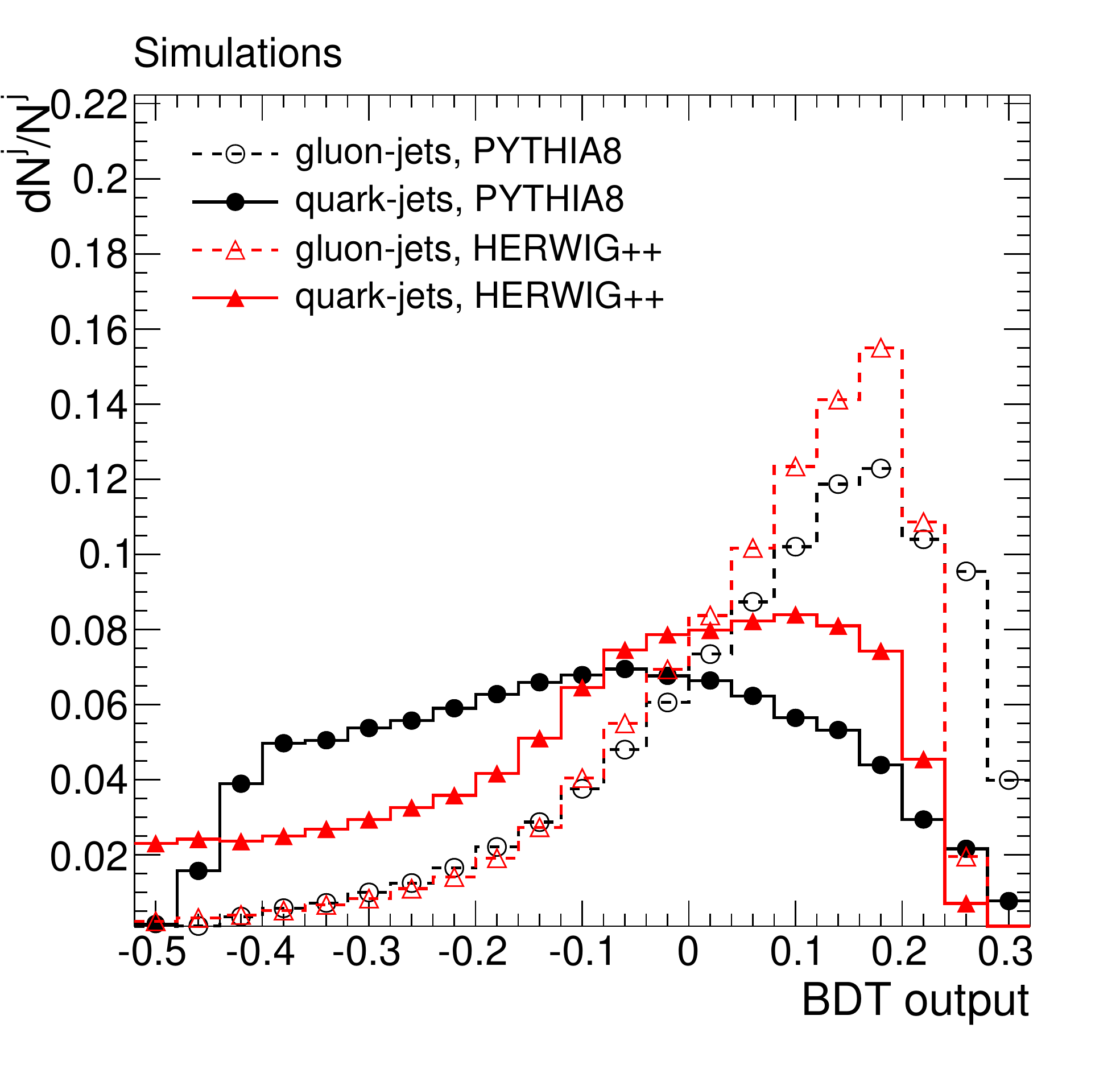}

\caption {{BDT output for gluon-initiated (hollow markers) jets and quark-initiated jets (solid markers) in case of  dijet events produced with \textsc{pythia}8 (black colored markers) and \textsc{herwig}++ (red colored markers). }} 
\label{bdt3}
\end{center}
\end{figure}

To further enhance the sensitivity and to consider the possible correlations between observables,
MVA method is used. This method is based on boosted decision trees (BDT) implemented in the TMVA 
framework~\cite{Hocker:2007ht}. The discriminating observables along with the $p_{\rm T}$ and 
$\eta$ of jets are provided as an input to the BDT. For MVA training, these observables are constructed using 
dijet events simulated with \textsc{pythia}8. To minimize the statistical 
bias, independent event samples of gluon- and quark-initiated jets are used for the training and 
testing of the MVA. The distribution of the BDT discriminant 
for gluon- and quark-initiated jets is shown in Fig.~\ref{bdt3}, which gives a clear 
distinction between two flavors of jets. The MVA output is used to tag a jet either as quark-initiated or gluon-initiated. The 
selected jets are tagged as gluon-initiated jets with requirement of BDT value greater than -0.105, 
otherwise considered as initiated by quarks. This criteria selects $\approx$ 90\% of gluon-initiated 
jets with 50\% rejection of quark-initiated jets. This trained MVA is used to select gluon-initiated jets in the Z + 2-jets events.
If both jets in the selected Z + 2-jets events are required to be initiated by gluons, SPS contribution got reduced by 48\% while keeping 72\% of the 
DPS events. In other words, DPS fraction in Z + 2-jets sample is now about 0.113 which is 20\% larger as compared to the cut-based analysis.

\begin{figure}[htbp]                                                                  
\begin{center}
\dimendef\prevdepth=0
\subfloat[]{\includegraphics[width=0.45\textwidth]{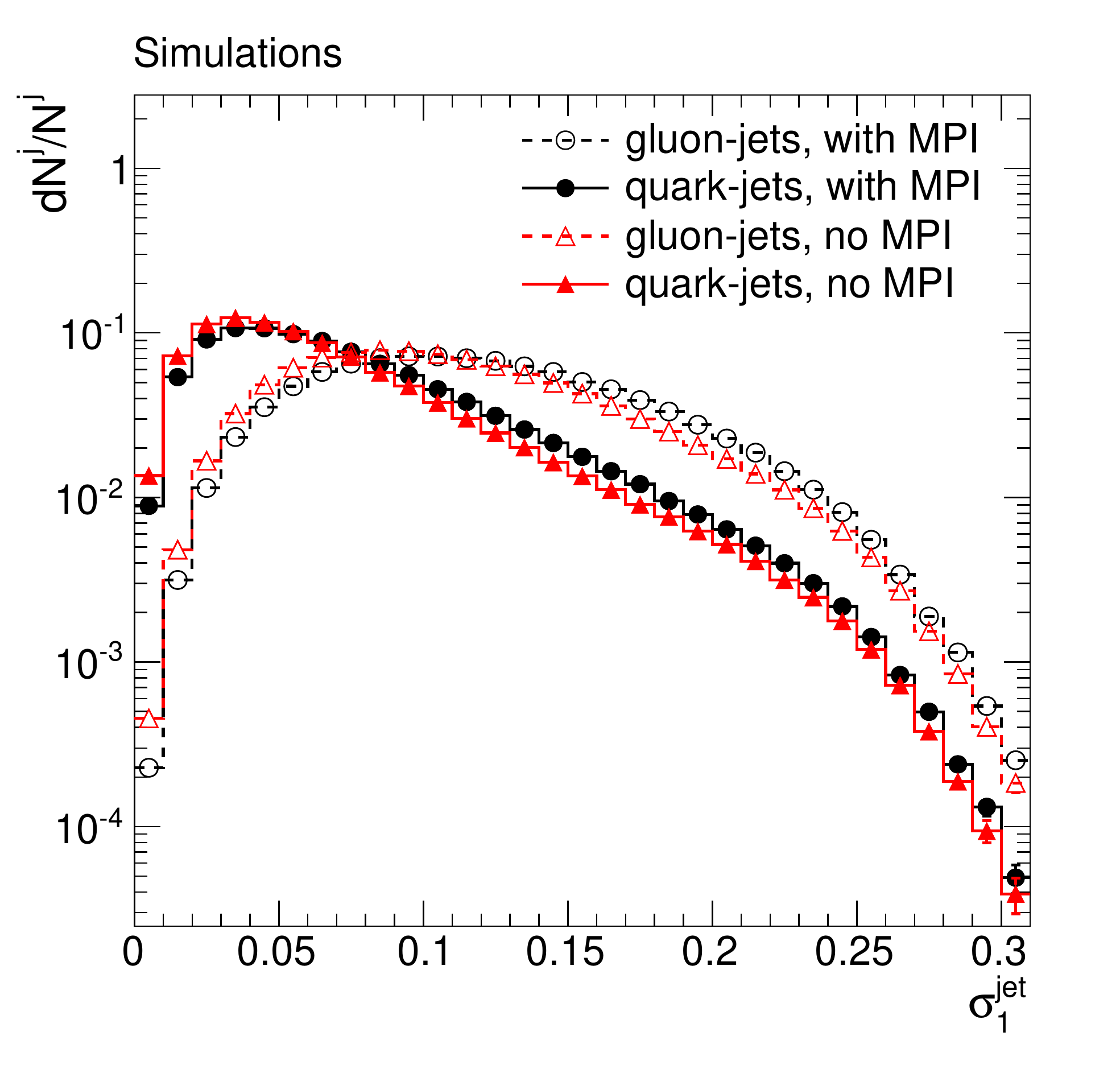}}
\subfloat[]{\includegraphics[width=0.45\textwidth]{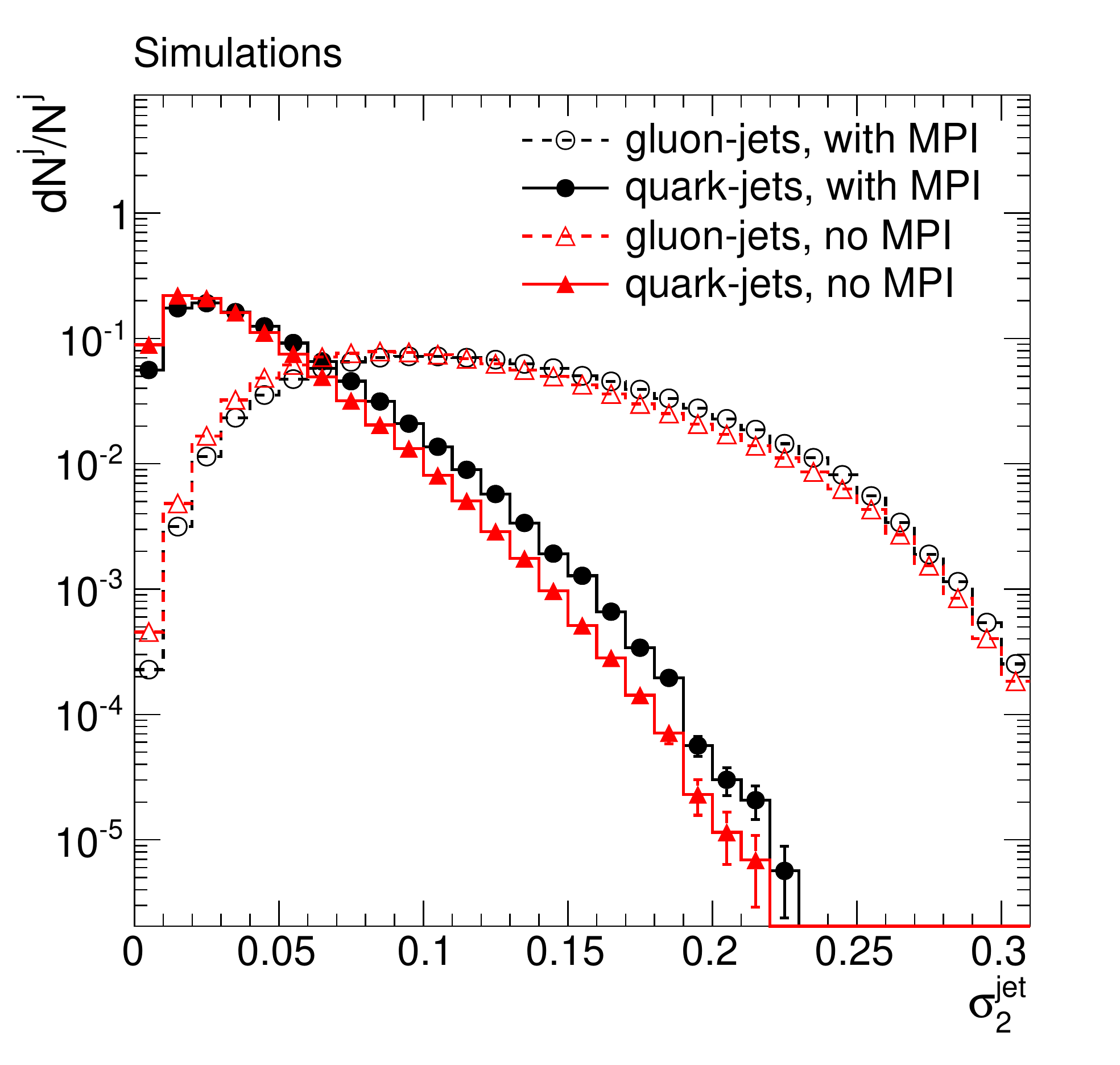}}\\
\subfloat[]{\includegraphics[width=0.45\textwidth]{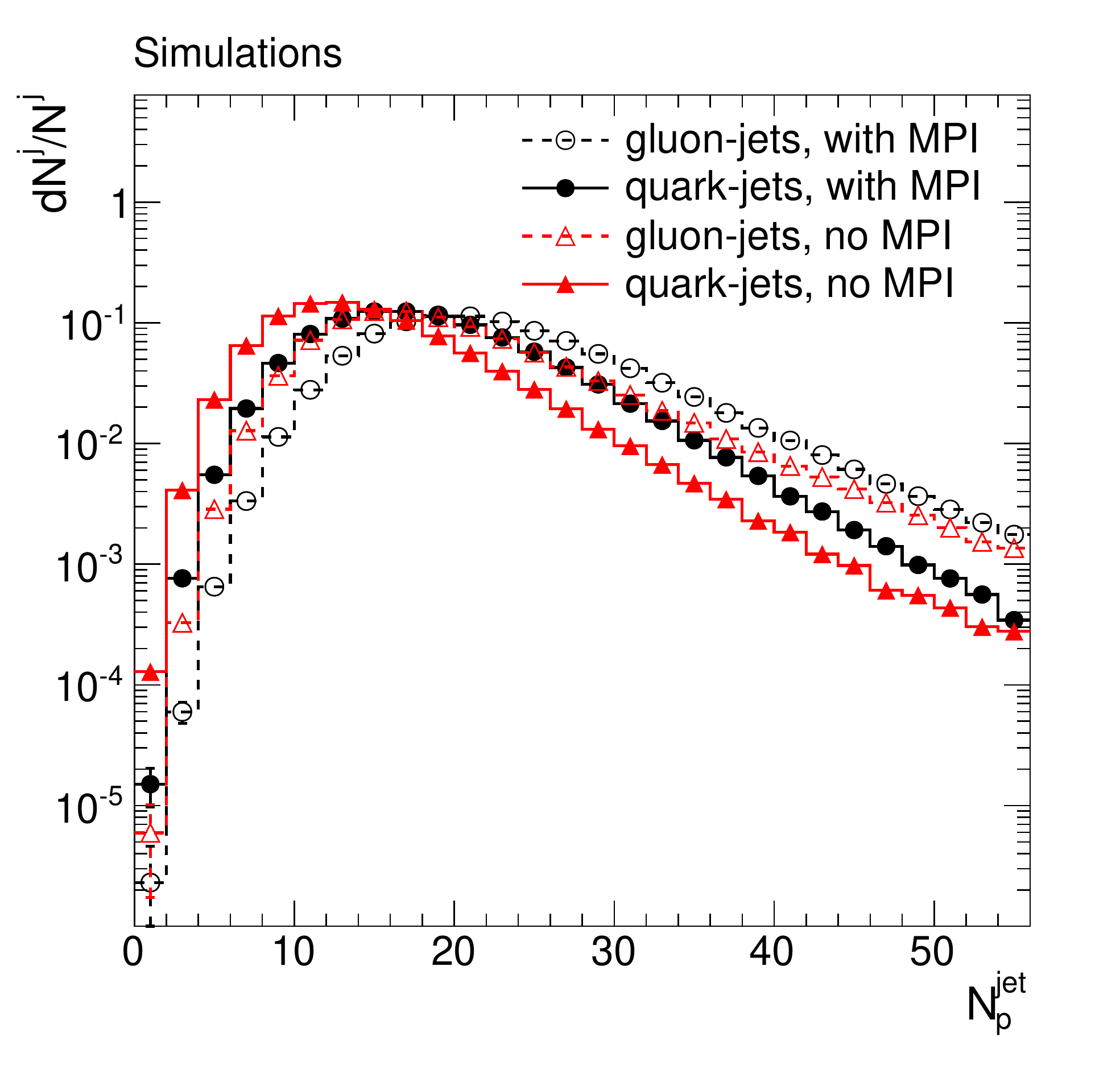}}
\subfloat[]{\includegraphics[width=0.45\textwidth]{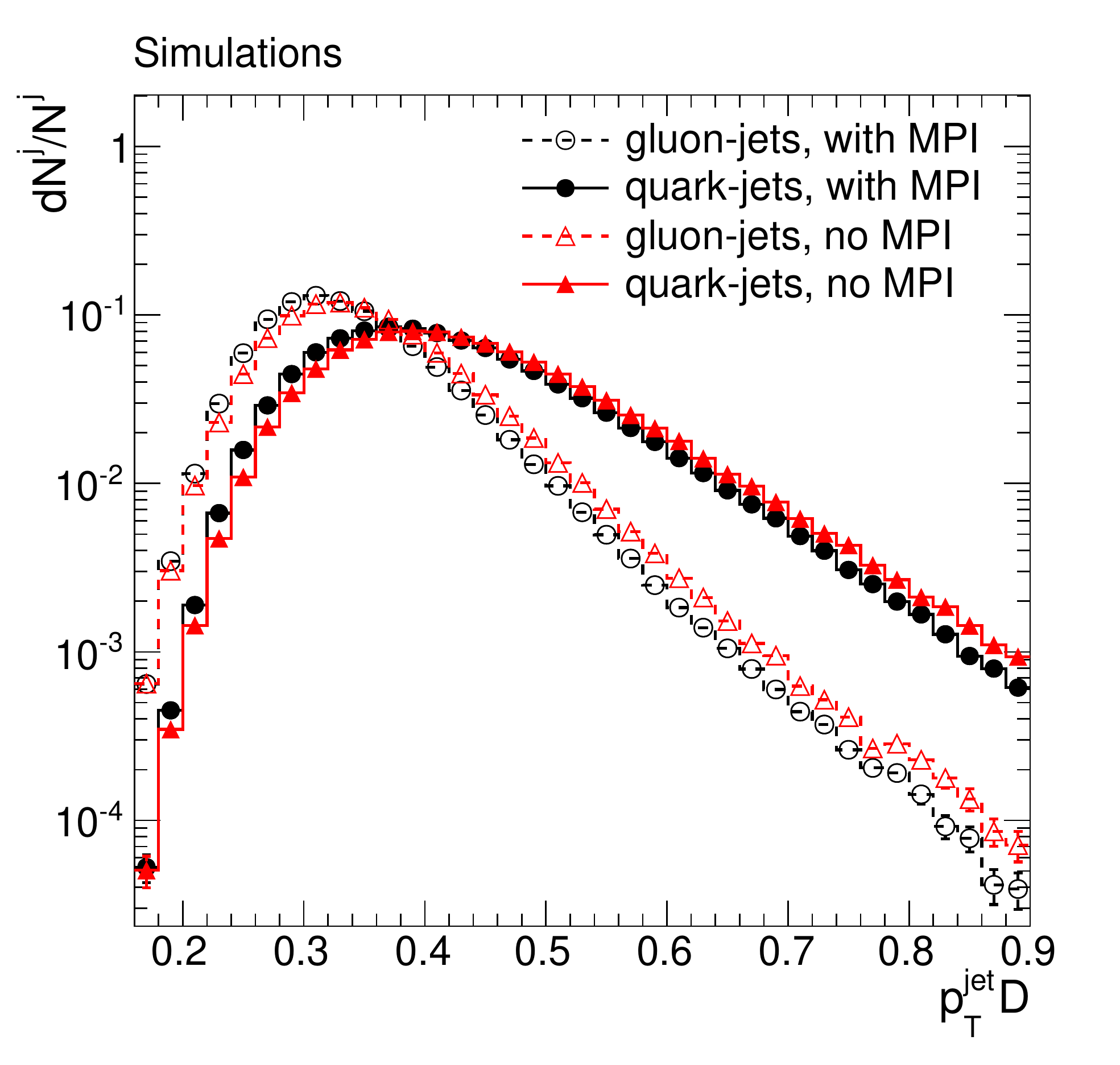}}

\caption {{The distributions of discriminating variables: 
(a) major axis size of jet cone ($\sigma_{1}^{\rm jet}$), (b) minor axis size of jet cone ($\sigma_{2}^{\rm jet}$), 
(c) jet constituents multiplicity ($N_{p}^{\rm jet}$), and (d) jet fragmentation function ($p_{\rm T}^{\rm jet}{D}$)
are compared for gluon-initiated jets (hollow markers) and quark-initiated jets (solid markers) Z + 2-jets events generated with \textsc{madgraph} + \textsc{pythia}8  with MPI (black colored markers) and without MPI (red colored markers).}} 
\label{comp_inpVarBDT2_mpi}
\end{center}
\end{figure}

The presence of MPI adds soft particles which can affect the intrinsic properties of a jet. It is 
observed that for the same quark-jet efficiency, the selection efficiency for gluon-initiated jets is about 92\% 
 if MPI is switched off during event generation. The effect of MPI on the input variables is shown in Fig.~\ref{comp_inpVarBDT2_mpi}.
 Different hadronization models can also affect the discrimination based on fragmentation properties of a jet. 
 As shown in Fig.~\ref{comp_inpVarBDT2}, there are significant differences in shape of input variables for \textsc{pythia}8 and \textsc{herwig}++.
 To investigate the possible effects of different hadronization models on the DPS fraction, the BDT discriminator is trained using events parton showered and hadronized with \textsc{herwig}++. 
 It is observed that for same efficiency of quark-initiated jets, the selection efficiency for gluon-initiated jets is 15\% lower if \textsc{herwig}++ is used  in place of \textsc{pythia}8 for the training of BDT. 
 This difference in hadronization properties results in reduction of DPS fraction to 0.107, but still there is a gain of 43\% as compared to selection when fragmentation properties of the jets were not used. 

To further corroborate effectiveness of 
the method, the Z + jets events are generated using \textsc{powheg}, which are hadronized 
and parton showered using \textsc{pythia}8. With this simulation, there is a similar 
rejection of SPS background, which results in 36\% gain in the DPS fraction for inclusive 
Z + 2-jets sample. The differences in the SPS rejection for \textsc{powheg} and \textsc{madgraph} 
is expected due to the differences in treatment of the LO and NLO effects which also change relative 
fraction of quark- and gluon-initiated jets. 

There are studies which show that SPS background can be 
suppressed by restricting the boost of Z-boson~\cite{Bansal:2016iri}. 
 The SPS background is suppressed with an upper cut on dilepton $p_{\rm T}$ less than 10 GeV/$c$ with DPS fraction of 0.32 in the selected events Z + 2-jets events.
The effect of jet-tagging is checked on top of the cut on the boost of Z-boson. Using jet fragmentation properties, in addition to Z-boost cut, as discriminator DPS fraction comes out to be 0.42. 
Now, gain in DPS fraction with jet fragmentation is relatively smaller as compared to no condition on the dilepton $p_{\rm T}$. 
This behavior is expected because after restricting the boost of Z-boson, most of the remaining jets will be coming from initial- or final-state radiation and hence, will be dominated by gluons.

The quark-gluon discriminator can be optimized for better discrimination between gluon- and quark-initiated jets with consideration of additional infrared and collinear (IRC) safe observables~\cite{Gallicchio:2011xq,Aad:2014gea,CMS:2013kfa,Cornelis:2014ima}. 
To study the effect of inclusion of IRC safe observables on the results, jet mass, jet shape and girth are considered in addition to the observables 
mentioned in Section~\ref{sec:methodology} to construct the discriminator. 
Qualitatively, the suppression of SPS background is still there and DPS fraction in selected Z + jets events increases to 0.119 , which is equivalent to  a gain of 59\% as compared to selection when fragmentation properties of the jets were not used.

It is conclusive from these studies that intrinsic properties of jets can be used to enhance the 
sensitivity to DPS, where two jets are produced from second interaction.  The different hadronization models 
have varying discrimination which can be considered as a part of the systematic uncertainty during the measurements.
The jets initiated by heavy quarks (in particular, b-quarks) show fragmentation properties similar 
to gluon-initiated jets. Therefore, it is required to veto these jets for effective separation 
between quark- and gluon-tagged jets. These jets, from b-quarks, can be efficiently identified 
with the use of available b-tagging algorithms~\cite{Aad:2015ydr,CMS:2016kkf}. The acceptance and pileup conditions in the LHC experiments can affect the above sensitivity and 
need to be studied with the actual experimental environment.

\section{Summary}\label{sec:summary}

This paper presents a feasibility study of the Z + jets events to explore the possibility to enhance DPS 
sensitivity by suppressing SPS background with the use of differences in the fragmentation 
properties of jets initiated by quarks and gluons. The Z + jets events are simulated with 
\textsc{madgraph} and \textsc{powheg} Monte-Carlo event generators, followed by 
hadronization and showering with \textsc{pythia}8. A number of observables, defining the 
fragmentation behavior of a jet, are used as input for the multivariate analysis and 
subsequently used to tag a jet as gluon-initiated or quark-initiated. It is observed 
that with the discrimination of gluon- and quark-initiated jet, it is possible to 
suppress SPS background up to 52\% resulting in a 40--50\% gain in Z + 2-jets 
events produced by DPS. 
The effect of different hadronization model on the discrimination is also investigated by using events simulated with \textsc{herwig}++.
As expected, \textsc{herwig}++ gives different sensitivity but SPS suppression is still there.
It will be interesting to study the impact of these variables 
on the DPS measurements in actual experimental conditions, $i.e.$, pileup, 
detector acceptance, at the LHC.

\begin{acknowledgments}
This work is financially supported by Department of Science and Technology (DST), 
New Delhi and University Grants Commission (UGC), New Delhi.

\end{acknowledgments}


\bibliography{paper_Zjets_PRD.bib}

\begin{thebibliography}{45}%
\makeatletter
\providecommand \@ifxundefined [1]{%
 \@ifx{#1\undefined}
}%
\providecommand \@ifnum [1]{%
 \ifnum #1\expandafter \@firstoftwo
 \else \expandafter \@secondoftwo
 \fi
}%
\providecommand \@ifx [1]{%
 \ifx #1\expandafter \@firstoftwo
 \else \expandafter \@secondoftwo
 \fi
}%
\providecommand \natexlab [1]{#1}%
\providecommand \enquote  [1]{``#1''}%
\providecommand \bibnamefont  [1]{#1}%
\providecommand \bibfnamefont [1]{#1}%
\providecommand \citenamefont [1]{#1}%
\providecommand \href@noop [0]{\@secondoftwo}%
\providecommand \href [0]{\begingroup \@sanitize@url \@href}%
\providecommand \@href[1]{\@@startlink{#1}\@@href}%
\providecommand \@@href[1]{\endgroup#1\@@endlink}%
\providecommand \@sanitize@url [0]{\catcode `\\12\catcode `\$12\catcode
  `\&12\catcode `\#12\catcode `\^12\catcode `\_12\catcode `\%12\relax}%
\providecommand \@@startlink[1]{}%
\providecommand \@@endlink[0]{}%
\providecommand \url  [0]{\begingroup\@sanitize@url \@url }%
\providecommand \@url [1]{\endgroup\@href {#1}{\urlprefix }}%
\providecommand \urlprefix  [0]{URL }%
\providecommand \Eprint [0]{\href }%
\providecommand \doibase [0]{http://dx.doi.org/}%
\providecommand \selectlanguage [0]{\@gobble}%
\providecommand \bibinfo  [0]{\@secondoftwo}%
\providecommand \bibfield  [0]{\@secondoftwo}%
\providecommand \translation [1]{[#1]}%
\providecommand \BibitemOpen [0]{}%
\providecommand \bibitemStop [0]{}%
\providecommand \bibitemNoStop [0]{.\EOS\space}%
\providecommand \EOS [0]{\spacefactor3000\relax}%
\providecommand \BibitemShut  [1]{\csname bibitem#1\endcsname}%
\let\auto@bib@innerbib\@empty
\bibitem [{\citenamefont {Sj{\"o}strand}\ and\ \citenamefont
  {Van~Zijl}(1987)}]{Sjostrand:1986ep}%
  \BibitemOpen
  \bibfield  {author} {\bibinfo {author} {\bibfnamefont {T.}~\bibnamefont
  {Sj{\"o}strand}}\ and\ \bibinfo {author} {\bibfnamefont {M.}~\bibnamefont
  {Van~Zijl}},\ }\href {\doibase 10.1103/PhysRevD.36.2019} {\bibfield
  {journal} {\bibinfo  {journal} {Phys. Rev. D}\ }\textbf {\bibinfo {volume}
  {36}},\ \bibinfo {pages} {2019} (\bibinfo {year} {1987})}\BibitemShut
  {NoStop}%
\bibitem [{\citenamefont {Diehl}\ \emph {et~al.}(2012)\citenamefont {Diehl},
  \citenamefont {Ostermeier},\ and\ \citenamefont
  {Sch{\"a}fer}}]{Diehl:2011yj}%
  \BibitemOpen
  \bibfield  {author} {\bibinfo {author} {\bibfnamefont {M.}~\bibnamefont
  {Diehl}}, \bibinfo {author} {\bibfnamefont {D.}~\bibnamefont {Ostermeier}}, \
  and\ \bibinfo {author} {\bibfnamefont {A.}~\bibnamefont {Sch{\"a}fer}},\
  }\href {\doibase 10.1007/JHEP03(2012)089} {\bibfield  {journal} {\bibinfo
  {journal} {JHEP}\ }\textbf {\bibinfo {volume} {03}},\ \bibinfo {pages} {89}
  (\bibinfo {year} {2012})},\ \Eprint {http://arxiv.org/abs/1111.0910}
  {arXiv:1111.0910 [hep-ph]} \BibitemShut {NoStop}%
\bibitem [{\citenamefont {Hussein}(2007)}]{Hussein:2006xr}%
  \BibitemOpen
  \bibfield  {author} {\bibinfo {author} {\bibfnamefont {M.~Y.}\ \bibnamefont
  {Hussein}},\ }\href {\doibase 10.1016/j.nuclphysbps.2007.08.086} {\bibfield
  {journal} {\bibinfo  {journal} {Nucl. Phys. Proc. Suppl.}\ }\textbf {\bibinfo
  {volume} {174}},\ \bibinfo {pages} {55} (\bibinfo {year} {2007})},\ \Eprint
  {http://arxiv.org/abs/hep-ph/0610207} {arXiv:hep-ph/0610207 [hep-ph]}
  \BibitemShut {NoStop}%
\bibitem [{\citenamefont {Chatrchyan}\ \emph {et~al.}(2011)\citenamefont
  {Chatrchyan} \emph {et~al.}}]{SUSY}%
  \BibitemOpen
  \bibfield  {author} {\bibinfo {author} {\bibnamefont {Chatrchyan}} \emph
  {et~al.} (\bibinfo {collaboration} {CMS}),\ }\href {\doibase
  10.1007/JHEP06(2011)077} {\bibfield  {journal} {\bibinfo  {journal} {JHEP}\
  }\textbf {\bibinfo {volume} {06}},\ \bibinfo {pages} {077} (\bibinfo {year}
  {2011})},\ \Eprint {http://arxiv.org/abs/1104.3168} {arXiv:1104.3168
  [hep-ex]} \BibitemShut {NoStop}%
\bibitem [{\citenamefont {Alitti}\ \emph {et~al.}(1991)\citenamefont {Alitti}
  \emph {et~al.}}]{jetua}%
  \BibitemOpen
  \bibfield  {author} {\bibinfo {author} {\bibfnamefont {J.}~\bibnamefont
  {Alitti}} \emph {et~al.} (\bibinfo {collaboration} {UA2}),\ }\href {\doibase
  10.1016/0370-2693(91)90937-L} {\bibfield  {journal} {\bibinfo  {journal}
  {Phys. Lett. B}\ }\textbf {\bibinfo {volume} {268}},\ \bibinfo {pages} {145}
  (\bibinfo {year} {1991})}\BibitemShut {NoStop}%
\bibitem [{\citenamefont {Akesson}\ \emph {et~al.}(1987)\citenamefont {Akesson}
  \emph {et~al.}}]{jetafs}%
  \BibitemOpen
  \bibfield  {author} {\bibinfo {author} {\bibfnamefont {T.}~\bibnamefont
  {Akesson}} \emph {et~al.} (\bibinfo {collaboration} {AFS}),\ }\href {\doibase
  10.1007/BF01566757} {\bibfield  {journal} {\bibinfo  {journal} {Z. Phys. C}\
  }\textbf {\bibinfo {volume} {34}},\ \bibinfo {pages} {163} (\bibinfo {year}
  {1987})}\BibitemShut {NoStop}%
\bibitem [{\citenamefont {Abazov}\ \emph {et~al.}(2010)\citenamefont {Abazov}
  \emph {et~al.}}]{photon3jetd0}%
  \BibitemOpen
  \bibfield  {author} {\bibinfo {author} {\bibfnamefont {V.~M.}\ \bibnamefont
  {Abazov}} \emph {et~al.} (\bibinfo {collaboration} {D0}),\ }\href {\doibase
  10.1103/PhysRevD.81.052012} {\bibfield  {journal} {\bibinfo  {journal} {Phys.
  Rev. D}\ }\textbf {\bibinfo {volume} {81}},\ \bibinfo {pages} {052012}
  (\bibinfo {year} {2010})},\ \Eprint {http://arxiv.org/abs/hep-ex/0912.5104}
  {arXiv:hep-ex/0912.5104 [hep-ex]} \BibitemShut {NoStop}%
\bibitem [{\citenamefont {Abe}\ \emph {et~al.}(1997)\citenamefont {Abe} \emph
  {et~al.}}]{photon3jetcdf}%
  \BibitemOpen
  \bibfield  {author} {\bibinfo {author} {\bibfnamefont {F.}~\bibnamefont
  {Abe}} \emph {et~al.} (\bibinfo {collaboration} {CDF}),\ }\href {\doibase
  10.1103/PhysRevD.56.3811} {\bibfield  {journal} {\bibinfo  {journal} {Phys.
  Rev. D}\ }\textbf {\bibinfo {volume} {56}},\ \bibinfo {pages} {3811}
  (\bibinfo {year} {1997})}\BibitemShut {NoStop}%
\bibitem [{\citenamefont {Abe}\ \emph {et~al.}(1993)\citenamefont {Abe} \emph
  {et~al.}}]{4jetcdf}%
  \BibitemOpen
  \bibfield  {author} {\bibinfo {author} {\bibfnamefont {F.}~\bibnamefont
  {Abe}} \emph {et~al.} (\bibinfo {collaboration} {CDF}),\ }\href {\doibase
  10.1103/PhysRevD.47.4857} {\bibfield  {journal} {\bibinfo  {journal} {Phys.
  Rev. D}\ }\textbf {\bibinfo {volume} {47}},\ \bibinfo {pages} {4857}
  (\bibinfo {year} {1993})}\BibitemShut {NoStop}%
\bibitem [{\citenamefont {Chatrchyan}\ \emph {et~al.}(2014)\citenamefont
  {Chatrchyan} \emph {et~al.}}]{w2jetcms}%
  \BibitemOpen
  \bibfield  {author} {\bibinfo {author} {\bibnamefont {Chatrchyan}} \emph
  {et~al.} (\bibinfo {collaboration} {CMS}),\ }\href {\doibase
  10.1007/JHEP03(2014)032} {\bibfield  {journal} {\bibinfo  {journal} {JHEP}\
  }\textbf {\bibinfo {volume} {03}},\ \bibinfo {pages} {032} (\bibinfo {year}
  {2014})},\ \Eprint {http://arxiv.org/abs/1312.5729} {arXiv:1312.5729
  [hep-ex]} \BibitemShut {NoStop}%
\bibitem [{\citenamefont {Aad}\ \emph {et~al.}(2013)\citenamefont {Aad} \emph
  {et~al.}}]{w2jetatlas}%
  \BibitemOpen
  \bibfield  {author} {\bibinfo {author} {\bibfnamefont {G.}~\bibnamefont
  {Aad}} \emph {et~al.} (\bibinfo {collaboration} {ATLAS}),\ }\href {\doibase
  10.1088/1367-2630/15/3/033038} {\bibfield  {journal} {\bibinfo  {journal}
  {New J. Phys.}\ }\textbf {\bibinfo {volume} {15}},\ \bibinfo {pages} {033038}
  (\bibinfo {year} {2013})},\ \Eprint {http://arxiv.org/abs/1301.6872}
  {arXiv:1301.6872 [hep-ex]} \BibitemShut {NoStop}%
\bibitem [{\citenamefont {Chatrchyan}\ \emph
  {et~al.}(2015{\natexlab{a}})\citenamefont {Chatrchyan} \emph
  {et~al.}}]{photon3jetcms}%
  \BibitemOpen
  \bibfield  {author} {\bibinfo {author} {\bibnamefont {Chatrchyan}} \emph
  {et~al.} (\bibinfo {collaboration} {CMS}),\ }\href@noop {} {\bibfield
  {journal} {\bibinfo  {journal} {CMS PAS}\ }\textbf {\bibinfo {volume}
  {FSQ-12-017}} (\bibinfo {year} {2015}{\natexlab{a}})}\BibitemShut {NoStop}%
\bibitem [{\citenamefont {Sirunyan}\ \emph {et~al.}(2018)\citenamefont
  {Sirunyan} \emph {et~al.}}]{dpswwcms8}%
  \BibitemOpen
  \bibfield  {author} {\bibinfo {author} {\bibfnamefont {A.~M.}\ \bibnamefont
  {Sirunyan}} \emph {et~al.} (\bibinfo {collaboration} {CMS}),\ }\href
  {\doibase 10.1007/JHEP02(2018)032} {\bibfield  {journal} {\bibinfo  {journal}
  {JHEP}\ }\textbf {\bibinfo {volume} {02}},\ \bibinfo {pages} {032} (\bibinfo
  {year} {2018})},\ \Eprint {http://arxiv.org/abs/1712.02280} {arXiv:1712.02280
  [hep-ex]} \BibitemShut {NoStop}%
\bibitem [{\citenamefont {Aaboud}\ \emph {et~al.}(2016)\citenamefont {Aaboud}
  \emph {et~al.}}]{Aaboud:2016dea}%
  \BibitemOpen
  \bibfield  {author} {\bibinfo {author} {\bibfnamefont {M.}~\bibnamefont
  {Aaboud}} \emph {et~al.} (\bibinfo {collaboration} {ATLAS}),\ }\href
  {\doibase 10.1007/JHEP11(2016)110} {\bibfield  {journal} {\bibinfo  {journal}
  {JHEP}\ }\textbf {\bibinfo {volume} {11}},\ \bibinfo {pages} {110} (\bibinfo
  {year} {2016})},\ \Eprint {http://arxiv.org/abs/1608.01857} {arXiv:1608.01857
  [hep-ex]} \BibitemShut {NoStop}%
\bibitem [{\citenamefont {Blok}\ and\ \citenamefont
  {Strikman}(2018)}]{Blok:2017alw}%
  \BibitemOpen
  \bibfield  {author} {\bibinfo {author} {\bibfnamefont {B.}~\bibnamefont
  {Blok}}\ and\ \bibinfo {author} {\bibfnamefont {M.}~\bibnamefont
  {Strikman}},\ }\href {\doibase 10.1142/9789813227767_0005} {\bibfield
  {journal} {\bibinfo  {journal} {Adv. Ser. Direct. High Energy Phys.}\
  }\textbf {\bibinfo {volume} {29}},\ \bibinfo {pages} {63} (\bibinfo {year}
  {2018})},\ \Eprint {http://arxiv.org/abs/1709.00334} {arXiv:1709.00334
  [hep-ph]} \BibitemShut {NoStop}%
\bibitem [{\citenamefont {Snigirev}(2010)}]{Snigirev:2010tk}%
  \BibitemOpen
  \bibfield  {author} {\bibinfo {author} {\bibfnamefont {A.~M.}\ \bibnamefont
  {Snigirev}},\ }\href {\doibase 10.1103/PhysRevD.81.065014} {\bibfield
  {journal} {\bibinfo  {journal} {Phys. Rev.}\ }\textbf {\bibinfo {volume}
  {D81}},\ \bibinfo {pages} {065014} (\bibinfo {year} {2010})},\ \Eprint
  {http://arxiv.org/abs/1001.0104} {arXiv:1001.0104 [hep-ph]} \BibitemShut
  {NoStop}%
\bibitem [{\citenamefont {Blok}\ and\ \citenamefont
  {Gunnellini}(2016)}]{Blok:2015afa}%
  \BibitemOpen
  \bibfield  {author} {\bibinfo {author} {\bibfnamefont {B.}~\bibnamefont
  {Blok}}\ and\ \bibinfo {author} {\bibfnamefont {P.}~\bibnamefont
  {Gunnellini}},\ }\href {\doibase 10.1140/epjc/s10052-016-4035-7} {\bibfield
  {journal} {\bibinfo  {journal} {Eur. Phys. J.}\ }\textbf {\bibinfo {volume}
  {C76}},\ \bibinfo {pages} {202} (\bibinfo {year} {2016})},\ \Eprint
  {http://arxiv.org/abs/1510.07436} {arXiv:1510.07436 [hep-ph]} \BibitemShut
  {NoStop}%
\bibitem [{\citenamefont {Maina}(2011)}]{Maina:2010vh}%
  \BibitemOpen
  \bibfield  {author} {\bibinfo {author} {\bibfnamefont {E.}~\bibnamefont
  {Maina}},\ }\href {\doibase 10.1007/JHEP01(2011)061} {\bibfield  {journal}
  {\bibinfo  {journal} {JHEP}\ }\textbf {\bibinfo {volume} {01}},\ \bibinfo
  {pages} {061} (\bibinfo {year} {2011})},\ \Eprint
  {http://arxiv.org/abs/1010.5674} {arXiv:1010.5674 [hep-ph]} \BibitemShut
  {NoStop}%
\bibitem [{\citenamefont {Cao}\ \emph {et~al.}(2018)\citenamefont {Cao},
  \citenamefont {Liu}, \citenamefont {Xie},\ and\ \citenamefont
  {Yan}}]{Cao:2017bcb}%
  \BibitemOpen
  \bibfield  {author} {\bibinfo {author} {\bibfnamefont {Q.-H.}\ \bibnamefont
  {Cao}}, \bibinfo {author} {\bibfnamefont {Y.}~\bibnamefont {Liu}}, \bibinfo
  {author} {\bibfnamefont {K.-P.}\ \bibnamefont {Xie}}, \ and\ \bibinfo
  {author} {\bibfnamefont {B.}~\bibnamefont {Yan}},\ }\href {\doibase
  10.1103/PhysRevD.97.035013} {\bibfield  {journal} {\bibinfo  {journal} {Phys.
  Rev.}\ }\textbf {\bibinfo {volume} {D97}},\ \bibinfo {pages} {035013}
  (\bibinfo {year} {2018})},\ \Eprint {http://arxiv.org/abs/1710.06315}
  {arXiv:1710.06315 [hep-ph]} \BibitemShut {NoStop}%
\bibitem [{\citenamefont {Alwall}\ \emph {et~al.}(2011)\citenamefont {Alwall}
  \emph {et~al.}}]{Alwall:2011}%
  \BibitemOpen
  \bibfield  {author} {\bibinfo {author} {\bibfnamefont {J.}~\bibnamefont
  {Alwall}} \emph {et~al.},\ }\href {\doibase 10.1007/JHEP06(2011)128}
  {\bibfield  {journal} {\bibinfo  {journal} {JHEP}\ }\textbf {\bibinfo
  {volume} {06}},\ \bibinfo {pages} {128} (\bibinfo {year} {2011})},\ \Eprint
  {http://arxiv.org/abs/1106.0522} {arXiv:1106.0522 [hep-ph]} \BibitemShut
  {NoStop}%
\bibitem [{\citenamefont {Maltoni}\ and\ \citenamefont
  {Stelzer}(2003)}]{Maltoni:2003}%
  \BibitemOpen
  \bibfield  {author} {\bibinfo {author} {\bibfnamefont {F.}~\bibnamefont
  {Maltoni}}\ and\ \bibinfo {author} {\bibfnamefont {T.}~\bibnamefont
  {Stelzer}},\ }\href {\doibase 10.1088/1126-6708/2003/02/027} {\bibfield
  {journal} {\bibinfo  {journal} {JHEP}\ }\textbf {\bibinfo {volume} {02}},\
  \bibinfo {pages} {027} (\bibinfo {year} {2003})},\ \Eprint
  {http://arxiv.org/abs/hep-ph/0208156} {arXiv:hep-ph/0208156 [hep-ph]}
  \BibitemShut {NoStop}%
\bibitem [{\citenamefont {Frixione}\ \emph {et~al.}(2007)\citenamefont
  {Frixione}, \citenamefont {Nason},\ and\ \citenamefont
  {Oleari}}]{Frixione:2007}%
  \BibitemOpen
  \bibfield  {author} {\bibinfo {author} {\bibfnamefont {S.}~\bibnamefont
  {Frixione}}, \bibinfo {author} {\bibfnamefont {P.}~\bibnamefont {Nason}}, \
  and\ \bibinfo {author} {\bibfnamefont {C.}~\bibnamefont {Oleari}},\ }\href
  {\doibase 10.1088/1126-6708/2007/11/070} {\bibfield  {journal} {\bibinfo
  {journal} {JHEP}\ }\textbf {\bibinfo {volume} {11}},\ \bibinfo {pages} {070}
  (\bibinfo {year} {2007})},\ \Eprint {http://arxiv.org/abs/0709.2092}
  {arXiv:0709.2092 [hep-ph]} \BibitemShut {NoStop}%
\bibitem [{\citenamefont {Campbell}\ \emph
  {et~al.}(2013{\natexlab{a}})\citenamefont {Campbell}, \citenamefont {Ellis},
  \citenamefont {Nason},\ and\ \citenamefont {Zanderighi}}]{PowhegW2J}%
  \BibitemOpen
  \bibfield  {author} {\bibinfo {author} {\bibfnamefont {J.~M.}\ \bibnamefont
  {Campbell}}, \bibinfo {author} {\bibfnamefont {R.~K.}\ \bibnamefont {Ellis}},
  \bibinfo {author} {\bibfnamefont {P.}~\bibnamefont {Nason}}, \ and\ \bibinfo
  {author} {\bibfnamefont {G.}~\bibnamefont {Zanderighi}},\ }\href {\doibase
  10.1007/JHEP08(2013)005} {\bibfield  {journal} {\bibinfo  {journal} {JHEP}\
  }\textbf {\bibinfo {volume} {08}},\ \bibinfo {pages} {005} (\bibinfo {year}
  {2013}{\natexlab{a}})},\ \Eprint {http://arxiv.org/abs/1303.5447}
  {arXiv:1303.5447 [hep-ph]} \BibitemShut {NoStop}%
\bibitem [{\citenamefont {Khachatryan}\ \emph {et~al.}(2017)\citenamefont
  {Khachatryan} \emph {et~al.}}]{Khachatryan:2016crw}%
  \BibitemOpen
  \bibfield  {author} {\bibinfo {author} {\bibfnamefont {V.}~\bibnamefont
  {Khachatryan}} \emph {et~al.} (\bibinfo {collaboration} {CMS}),\ }\href
  {\doibase 10.1007/JHEP04(2017)022} {\bibfield  {journal} {\bibinfo  {journal}
  {JHEP}\ }\textbf {\bibinfo {volume} {04}},\ \bibinfo {pages} {022} (\bibinfo
  {year} {2017})},\ \Eprint {http://arxiv.org/abs/1611.03844} {arXiv:1611.03844
  [hep-ex]} \BibitemShut {NoStop}%
\bibitem [{\citenamefont {Hamilton}\ \emph {et~al.}(2012)\citenamefont
  {Hamilton}, \citenamefont {Nason},\ and\ \citenamefont {Zanderighi}}]{MINLO}%
  \BibitemOpen
  \bibfield  {author} {\bibinfo {author} {\bibfnamefont {K.}~\bibnamefont
  {Hamilton}}, \bibinfo {author} {\bibfnamefont {P.}~\bibnamefont {Nason}}, \
  and\ \bibinfo {author} {\bibfnamefont {G.}~\bibnamefont {Zanderighi}},\
  }\href {\doibase 10.1007/JHEP10(2012)155} {\bibfield  {journal} {\bibinfo
  {journal} {JHEP}\ }\textbf {\bibinfo {volume} {10}},\ \bibinfo {pages} {155}
  (\bibinfo {year} {2012})},\ \Eprint {http://arxiv.org/abs/1206.3572}
  {arXiv:1206.3572 [hep-ph]} \BibitemShut {NoStop}%
\bibitem [{\citenamefont {Campbell}\ \emph
  {et~al.}(2013{\natexlab{b}})\citenamefont {Campbell}, \citenamefont {Ellis},
  \citenamefont {Nason},\ and\ \citenamefont {Zanderighi}}]{Campbell:2013vha}%
  \BibitemOpen
  \bibfield  {author} {\bibinfo {author} {\bibfnamefont {J.~M.}\ \bibnamefont
  {Campbell}}, \bibinfo {author} {\bibfnamefont {R.~K.}\ \bibnamefont {Ellis}},
  \bibinfo {author} {\bibfnamefont {P.}~\bibnamefont {Nason}}, \ and\ \bibinfo
  {author} {\bibfnamefont {G.}~\bibnamefont {Zanderighi}},\ }\href {\doibase
  10.1007/JHEP08(2013)005} {\bibfield  {journal} {\bibinfo  {journal} {JHEP}\
  }\textbf {\bibinfo {volume} {08}},\ \bibinfo {pages} {005} (\bibinfo {year}
  {2013}{\natexlab{b}})},\ \Eprint {http://arxiv.org/abs/1303.5447}
  {arXiv:1303.5447 [hep-ph]} \BibitemShut {NoStop}%
\bibitem [{\citenamefont {Sj{\"o}strand}\ \emph {et~al.}(2008)\citenamefont
  {Sj{\"o}strand}, \citenamefont {Mrenna},\ and\ \citenamefont
  {Skands}}]{Sjostrand:2007gs}%
  \BibitemOpen
  \bibfield  {author} {\bibinfo {author} {\bibfnamefont {T.}~\bibnamefont
  {Sj{\"o}strand}}, \bibinfo {author} {\bibfnamefont {S.}~\bibnamefont
  {Mrenna}}, \ and\ \bibinfo {author} {\bibfnamefont {P.~Z.}\ \bibnamefont
  {Skands}},\ }\href {\doibase 10.1016/j.cpc.2008.01.036} {\bibfield  {journal}
  {\bibinfo  {journal} {Comput. Phys. Commun.}\ }\textbf {\bibinfo {volume}
  {178}},\ \bibinfo {pages} {852} (\bibinfo {year} {2008})},\ \Eprint
  {http://arxiv.org/abs/0710.3820} {arXiv:0710.3820 [hep-ph]} \BibitemShut
  {NoStop}%
\bibitem [{\citenamefont {Corke}(2009)}]{Corke:2009pm}%
  \BibitemOpen
  \bibfield  {author} {\bibinfo {author} {\bibfnamefont {R.}~\bibnamefont
  {Corke}},\ }\href@noop {} {\  (\bibinfo {year} {2009})},\ \Eprint
  {http://arxiv.org/abs/0901.2852} {arXiv:0901.2852 [hep-ph]} \BibitemShut
  {NoStop}%
\bibitem [{\citenamefont {Aad}\ \emph {et~al.}(2012)\citenamefont {Aad} \emph
  {et~al.}}]{ATLAS:2012uec}%
  \BibitemOpen
  \bibfield  {author} {\bibinfo {author} {\bibfnamefont {G.}~\bibnamefont
  {Aad}} \emph {et~al.} (\bibinfo {collaboration} {ATLAS}),\ }\href@noop {}
  {\bibfield  {journal} {\bibinfo  {journal} {ATL-PHYS}\ }\textbf {\bibinfo
  {volume} {PUB-2012-003}} (\bibinfo {year} {2012})}\BibitemShut {NoStop}%
\bibitem [{\citenamefont {Bahr}\ \emph {et~al.}(2008)\citenamefont {Bahr} \emph
  {et~al.}}]{Bahr:2008pv}%
  \BibitemOpen
  \bibfield  {author} {\bibinfo {author} {\bibfnamefont {M.}~\bibnamefont
  {Bahr}} \emph {et~al.},\ }\href {\doibase 10.1140/epjc/s10052-008-0798-9}
  {\bibfield  {journal} {\bibinfo  {journal} {Eur. Phys. J.}\ }\textbf
  {\bibinfo {volume} {C58}},\ \bibinfo {pages} {639} (\bibinfo {year}
  {2008})},\ \Eprint {http://arxiv.org/abs/0803.0883} {arXiv:0803.0883
  [hep-ph]} \BibitemShut {NoStop}%
\bibitem [{\citenamefont {Chatrchyan}\ \emph
  {et~al.}(2015{\natexlab{b}})\citenamefont {Chatrchyan} \emph
  {et~al.}}]{Khachatryan:2015pea}%
  \BibitemOpen
  \bibfield  {author} {\bibinfo {author} {\bibnamefont {Chatrchyan}} \emph
  {et~al.} (\bibinfo {collaboration} {CMS}),\ }\href@noop {} {\  (\bibinfo
  {year} {2015}{\natexlab{b}})},\ \Eprint {http://arxiv.org/abs/1512.00815}
  {arXiv:1512.00815 [hep-ex]} \BibitemShut {NoStop}%
\bibitem [{\citenamefont {Cacciari}\ \emph {et~al.}(2008)\citenamefont
  {Cacciari}, \citenamefont {Salam},\ and\ \citenamefont
  {Soyez}}]{Cacciari:2008gp}%
  \BibitemOpen
  \bibfield  {author} {\bibinfo {author} {\bibfnamefont {M.}~\bibnamefont
  {Cacciari}}, \bibinfo {author} {\bibfnamefont {G.~P.}\ \bibnamefont {Salam}},
  \ and\ \bibinfo {author} {\bibfnamefont {G.}~\bibnamefont {Soyez}},\ }\href
  {\doibase 10.1088/1126-6708/2008/04/063} {\bibfield  {journal} {\bibinfo
  {journal} {JHEP}\ }\textbf {\bibinfo {volume} {04}},\ \bibinfo {pages} {063}
  (\bibinfo {year} {2008})},\ \Eprint {http://arxiv.org/abs/0802.1189}
  {arXiv:0802.1189 [hep-ph]} \BibitemShut {NoStop}%
\bibitem [{\citenamefont {Cacciari}\ \emph {et~al.}(2012)\citenamefont
  {Cacciari}, \citenamefont {Salam},\ and\ \citenamefont
  {Soyez}}]{Cacciari:2011ma}%
  \BibitemOpen
  \bibfield  {author} {\bibinfo {author} {\bibfnamefont {M.}~\bibnamefont
  {Cacciari}}, \bibinfo {author} {\bibfnamefont {G.~P.}\ \bibnamefont {Salam}},
  \ and\ \bibinfo {author} {\bibfnamefont {G.}~\bibnamefont {Soyez}},\ }\href
  {\doibase 10.1140/epjc/s10052-012-1896-2} {\bibfield  {journal} {\bibinfo
  {journal} {Eur. Phys. J.}\ }\textbf {\bibinfo {volume} {C72}},\ \bibinfo
  {pages} {1896} (\bibinfo {year} {2012})},\ \Eprint
  {http://arxiv.org/abs/1111.6097} {arXiv:1111.6097 [hep-ph]} \BibitemShut
  {NoStop}%
\bibitem [{\citenamefont {Collaboration}(1993)}]{OPAL1993:qg1}%
  \BibitemOpen
  \bibfield  {author} {\bibinfo {author} {\bibfnamefont {O.}~\bibnamefont
  {Collaboration}},\ }\href {\doibase 10.1007/BF01557696} {\bibfield  {journal}
  {\bibinfo  {journal} {Zeitschrift f{\"u}r Physik C Particles and Fields}\
  }\textbf {\bibinfo {volume} {58}},\ \bibinfo {pages} {387} (\bibinfo {year}
  {1993})}\BibitemShut {NoStop}%
\bibitem [{\citenamefont {Akers}\ \emph {et~al.}(1995)\citenamefont {Akers}
  \emph {et~al.}}]{OPAL:1995ab}%
  \BibitemOpen
  \bibfield  {author} {\bibinfo {author} {\bibfnamefont {R.}~\bibnamefont
  {Akers}} \emph {et~al.} (\bibinfo {collaboration} {OPAL}),\ }\href@noop {}
  {\bibfield  {journal} {\bibinfo  {journal} {Z. Phys.}\ }\textbf {\bibinfo
  {volume} {C68}},\ \bibinfo {pages} {179} (\bibinfo {year}
  {1995})}\BibitemShut {NoStop}%
\bibitem [{\citenamefont {Abreu}\ \emph {et~al.}(1996)\citenamefont {Abreu}
  \emph {et~al.}}]{Abreu:1995hp}%
  \BibitemOpen
  \bibfield  {author} {\bibinfo {author} {\bibfnamefont {P.}~\bibnamefont
  {Abreu}} \emph {et~al.} (\bibinfo {collaboration} {DELPHI}),\ }\href
  {\doibase 10.1007/s002880050095} {\bibfield  {journal} {\bibinfo  {journal}
  {Z. Phys.}\ }\textbf {\bibinfo {volume} {C70}},\ \bibinfo {pages} {179}
  (\bibinfo {year} {1996})}\BibitemShut {NoStop}%
\bibitem [{\citenamefont {Buskulic}\ \emph {et~al.}(1996)\citenamefont
  {Buskulic} \emph {et~al.}}]{Buskulic:1995sw}%
  \BibitemOpen
  \bibfield  {author} {\bibinfo {author} {\bibfnamefont {D.}~\bibnamefont
  {Buskulic}} \emph {et~al.} (\bibinfo {collaboration} {ALEPH}),\ }\href
  {\doibase 10.1016/0370-2693(96)00849-0} {\bibfield  {journal} {\bibinfo
  {journal} {Phys. Lett.}\ }\textbf {\bibinfo {volume} {B384}},\ \bibinfo
  {pages} {353} (\bibinfo {year} {1996})}\BibitemShut {NoStop}%
\bibitem [{\citenamefont {Gallicchio}\ and\ \citenamefont
  {Schwartz}(2011)}]{Gallicchio:2011xq}%
  \BibitemOpen
  \bibfield  {author} {\bibinfo {author} {\bibfnamefont {J.}~\bibnamefont
  {Gallicchio}}\ and\ \bibinfo {author} {\bibfnamefont {M.~D.}\ \bibnamefont
  {Schwartz}},\ }\href {\doibase 10.1103/PhysRevLett.107.172001} {\bibfield
  {journal} {\bibinfo  {journal} {Phys. Rev. Lett.}\ }\textbf {\bibinfo
  {volume} {107}},\ \bibinfo {pages} {172001} (\bibinfo {year} {2011})},\
  \Eprint {http://arxiv.org/abs/1106.3076} {arXiv:1106.3076 [hep-ph]}
  \BibitemShut {NoStop}%
\bibitem [{\citenamefont {Aad}\ \emph {et~al.}(2014)\citenamefont {Aad} \emph
  {et~al.}}]{Aad:2014gea}%
  \BibitemOpen
  \bibfield  {author} {\bibinfo {author} {\bibfnamefont {G.}~\bibnamefont
  {Aad}} \emph {et~al.} (\bibinfo {collaboration} {ATLAS}),\ }\href {\doibase
  10.1140/epjc/s10052-014-3023-z} {\bibfield  {journal} {\bibinfo  {journal}
  {Eur. Phys. J.}\ }\textbf {\bibinfo {volume} {C74}},\ \bibinfo {pages} {3023}
  (\bibinfo {year} {2014})},\ \Eprint {http://arxiv.org/abs/1405.6583}
  {arXiv:1405.6583 [hep-ex]} \BibitemShut {NoStop}%
\bibitem [{\citenamefont {Chatrchyan}\ \emph {et~al.}(2013)\citenamefont
  {Chatrchyan} \emph {et~al.}}]{CMS:2013kfa}%
  \BibitemOpen
  \bibfield  {author} {\bibinfo {author} {\bibnamefont {Chatrchyan}} \emph
  {et~al.} (\bibinfo {collaboration} {CMS}),\ }\href@noop {} {\bibfield
  {journal} {\bibinfo  {journal} {CMS PAS}\ }\textbf {\bibinfo {volume}
  {JME-13-002}} (\bibinfo {year} {2013})}\BibitemShut {NoStop}%
\bibitem [{\citenamefont {Cornelis}(2014)}]{Cornelis:2014ima}%
  \BibitemOpen
  \bibfield  {author} {\bibinfo {author} {\bibfnamefont {T.}~\bibnamefont
  {Cornelis}} (\bibinfo {collaboration} {CMS}),\ }in\ \href
  {https://inspirehep.net/record/1315816/files/arXiv:1409.3072.pdf} {\emph
  {\bibinfo {booktitle} {{Proceedings, 2nd Conference on Large Hadron Collider
  Physics Conference (LHCP 2014): New York, USA, June 2-7, 2014}}}}\ (\bibinfo
  {year} {2014})\ \Eprint {http://arxiv.org/abs/1409.3072} {arXiv:1409.3072
  [hep-ex]} \BibitemShut {NoStop}%
\bibitem [{\citenamefont {Hoecker}\ \emph {et~al.}(2007)\citenamefont
  {Hoecker}, \citenamefont {Speckmayer}, \citenamefont {Stelzer}, \citenamefont
  {Therhaag}, \citenamefont {von Toerne},\ and\ \citenamefont
  {Voss}}]{Hocker:2007ht}%
  \BibitemOpen
  \bibfield  {author} {\bibinfo {author} {\bibfnamefont {A.}~\bibnamefont
  {Hoecker}}, \bibinfo {author} {\bibfnamefont {P.}~\bibnamefont {Speckmayer}},
  \bibinfo {author} {\bibfnamefont {J.}~\bibnamefont {Stelzer}}, \bibinfo
  {author} {\bibfnamefont {J.}~\bibnamefont {Therhaag}}, \bibinfo {author}
  {\bibfnamefont {E.}~\bibnamefont {von Toerne}}, \ and\ \bibinfo {author}
  {\bibfnamefont {H.}~\bibnamefont {Voss}},\ }\href@noop {} {\bibfield
  {journal} {\bibinfo  {journal} {PoS}\ }\textbf {\bibinfo {volume} {ACAT}},\
  \bibinfo {pages} {040} (\bibinfo {year} {2007})},\ \Eprint
  {http://arxiv.org/abs/physics/0703039} {arXiv:physics/0703039} \BibitemShut
  {NoStop}%
\bibitem [{\citenamefont {Kumar}\ \emph {et~al.}(2016)\citenamefont {Kumar},
  \citenamefont {Bansal}, \citenamefont {Bansal},\ and\ \citenamefont
  {Singh}}]{Bansal:2016iri}%
  \BibitemOpen
  \bibfield  {author} {\bibinfo {author} {\bibfnamefont {R.}~\bibnamefont
  {Kumar}}, \bibinfo {author} {\bibfnamefont {M.}~\bibnamefont {Bansal}},
  \bibinfo {author} {\bibfnamefont {S.}~\bibnamefont {Bansal}}, \ and\ \bibinfo
  {author} {\bibfnamefont {J.~B.}\ \bibnamefont {Singh}},\ }\href {\doibase
  10.1103/PhysRevD.93.054019} {\bibfield  {journal} {\bibinfo  {journal} {Phys.
  Rev.}\ }\textbf {\bibinfo {volume} {D93}},\ \bibinfo {pages} {054019}
  (\bibinfo {year} {2016})},\ \Eprint {http://arxiv.org/abs/1602.05392}
  {arXiv:1602.05392 [hep-ph]} \BibitemShut {NoStop}%
\bibitem [{\citenamefont {Aad}\ \emph {et~al.}(2016)\citenamefont {Aad} \emph
  {et~al.}}]{Aad:2015ydr}%
  \BibitemOpen
  \bibfield  {author} {\bibinfo {author} {\bibfnamefont {G.}~\bibnamefont
  {Aad}} \emph {et~al.} (\bibinfo {collaboration} {ATLAS}),\ }\href {\doibase
  10.1088/1748-0221/11/04/P04008} {\bibfield  {journal} {\bibinfo  {journal}
  {JINST}\ }\textbf {\bibinfo {volume} {11}},\ \bibinfo {pages} {P04008}
  (\bibinfo {year} {2016})},\ \Eprint {http://arxiv.org/abs/1512.01094}
  {arXiv:1512.01094 [hep-ex]} \BibitemShut {NoStop}%
\bibitem [{\citenamefont {Chatrchyan}\ \emph {et~al.}(2016)\citenamefont
  {Chatrchyan} \emph {et~al.}}]{CMS:2016kkf}%
  \BibitemOpen
  \bibfield  {author} {\bibinfo {author} {\bibnamefont {Chatrchyan}} \emph
  {et~al.} (\bibinfo {collaboration} {CMS}),\ }\href@noop {} {\bibfield
  {journal} {\bibinfo  {journal} {CMS PAS}\ }\textbf {\bibinfo {volume}
  {BTV-15-001}} (\bibinfo {year} {2016})}\BibitemShut {NoStop}%
\end{thebibliography}%

\end{document}